\documentclass{article}

\usepackage{graphicx}
\usepackage{amsthm}
\usepackage[centertags]{amsmath}
\usepackage{amsgen}
\usepackage{amscd}
\usepackage{amsbsy}
\usepackage{amsopn}
\usepackage{amstext}
\usepackage{amsxtra}
\usepackage{amsfonts}
\usepackage{epsfig}
\usepackage[a4paper]{geometry} %% This package takes care of the margins in a more efficient way than LaTeX

\usepackage[english]{babel}

\newcommand{\bear}{\begin{array}}
\newcommand{\ear}{\end{array}}

\newcommand{\beq}{\begin{equation}}
\newcommand{\eeq}{\end{equation}}
\newcommand{\beqa}{\begin{eqnarray}}
\newcommand{\eeqa}{\end{eqnarray}}

\newcommand{\nn}{\nonumber}
\def\OMIT#1{{}}
\newcommand{\lsim}{\mathrel{\rlap{\lower4pt\hbox{\hskip1pt$\sim$}}
    \raise1pt\hbox{$<$}}}         %less than or approx. symbol
\newcommand{\gsim}{\mathrel{\rlap{\lower4pt\hbox{\hskip1pt$\sim$}}
    \raise1pt\hbox{$>$}}}         %greater than or approx. symbol

\newcommand{\tc}{t_{\rm c}}
\newcommand{\te}{t_{\rm esc}}
\newcommand{\R}{\mathcal{R}}
\newcommand{\X}{X_{\rm esc}}

\begin{document}

\vskip1.5cm
\begin{center}
  {\Large \bf Cosmic ray propagation time scales: lessons from radioactive nuclei and positron data}\\
\end{center}
\vskip0.2cm

\begin{center}
{\bf Kfir Blum}

\end{center}
\vskip 8pt

\begin{center}
{\it Department of Particle Physics and Astrophysics\\
Weizmann Institute of Science,\\ Rehovot 76100, Israel} \vspace*{0.3cm}

{\tt  kfir.blum@weizmann.ac.il}
\end{center}

\vglue 0.3truecm

\begin{abstract}
We take a fresh look at high energy radioactive nuclei data reported in the 90's and at the positron data recently reported by PAMELA. Our aim is to study the model independent implications of these data for the propagation time scales of cosmic rays in the Galaxy.
Considering radioactive nuclei, using decaying charge to decayed charge ratios -- the only directly relevant data available at relativistic energies -- we show that a rigidity independent residence time is consistent with observations. The data for all nuclei can be described by $f_{s,i}=\left(t_i/100\,{\rm Myr}\right)^{0.7}$, where $f_{s,i}$ is the suppression of the flux due to decay and $t_i$ is the observer frame lifetime for nucleus specie $i$.
Considering positron measurements, we argue that the positron flux is consistent with a secondary origin. Comparing the positron data with radioactive nuclei at the same energy range, we derive an upper bound on the mean electromagnetic energy density traversed by the positrons, $\bar U_T<1.25$ eV/cm$^3$ at a rigidity of $\R=$ 40 GV. Charge ratio measurements within easy reach of the AMS-02 experiment, most notably a determination of the Cl/Ar ratio extending up to $\R\sim$ 100 GV, will constrain the energy dependence of the positron cooling time. Such constraints can be used to distinguish between different propagation scenarios, as well as to test the secondary origin hypothesis for the positrons in detail.
\end{abstract}

\section{Introduction}\label{s:int}

The sources of primary cosmic rays (CRs), the correct description of CR propagation and the mechanisms of CR trapping in and escape from the Galaxy are all essentially unknown~\cite{ginbook,kulbook,longbook}. Given the list of open questions, it is remarkable that very few detailed model independent analyses exist in the literature. In contrast, models of homogeneous diffusion have become a standard, where the general approach is to fit the parameters of the model to B/C and low energy $^{10}$Be/$^9$Be data. It is difficult to extract generally applicable information from such studies.

In this work we present a model independent analysis of the propagation time scales of CR nuclei and positrons. These time scales are related to the CR residence time in the Galaxy, a key unknown of CR propagation. Motivated by the prospects for improved measurements in the near future~\cite{Adriani:2008zr,Ahn:2008my,collaboration:2010ij,Guzik:2008zz,Bindi:2010zz}, our analysis demonstrates that it is possible to extract significant quantitative information from CR measurements under general assumptions, without committing to any particular propagation model.

The main data we use are based on the decaying charge to decayed charge ratios Be/B, Al/Mg and Cl/Ar~\cite{SS70}, measured to relativistic energies~\cite{Engelmann:1990zz} and studied more than a decade ago by~\cite{WS98,Ptuskin1999}. Earlier analyses of CR propagation time scales in the context of diffusion and leaky box models can be found e.g. in~\cite{SGM88,Ptuskin:1998,Yanasak:2001,Mewaldt:2001,Donato:2001eq,Strong:2007nh} and references therein. Being model dependent, the values deduced in these studies for the CR residence time differ by order of magnitude, as well as by their physical interpretation. In addition, these studies focused on low energy isotopic data, where various theoretical complications arise and where the limited dynamical range of the experiments precludes a direct inference of the energy dependence of the residence time. Apart from the work of~\cite{WS98,Ptuskin1999}, the high energy charge ratio measurements we analyze here were recently considered in~\cite{Shibata,Putze:2010zn} in the context of fits to the parameters of diffusion models.

Before proceeding to describe the plan of this paper we first emphasize the key concepts and approximations. A central tool in this work is the CR grammage, the mean traversed ISM column density experienced by CRs~\cite{ginbook}. At high energy, where propagation energy gains/losses can be neglected, the local flux of any stable secondary nucleus can be calculated reliably as~\cite{W97,Katz:2009yd}
\beq\label{eq:Xeq} J_S=\frac{c}{4\pi}\,\X\,\tilde Q_S,\eeq
where $\X$ is the CR grammage and $\tilde Q_S$ is the local net production (including spallation losses) per unit column density of ISM.  Eq.~(\ref{eq:Xeq}) is an empirical relation; any acceptable model of CR propagation must reproduce it.

In our study we identify the effects due to decay in radioactive nuclei and due to radiative losses in positrons, by using Eq.~(\ref{eq:Xeq}) to compute the flux that would be expected had there been no decay and loss and relating it to the flux observed in practice. Doing so entails some approximation.
For Eq.~(\ref{eq:Xeq}) to apply, the CR rigidity should remain unchanged during propagation, including fragmentation and decay events. In contrast, the decaying charge and decayed charge samples, reported at a given kinetic energy per nucleon, exhibit a spread in rigidity. This spread is more pronounced for lighter elements, the maximal change occurring in the decay $^{10}_4{\rm Be}\to ^{10}_5$B in which the rigidity of the nucleus drops by 20\%. Similar effects occur in certain spallation reactions and in the production of positrons and antiprotons, which do not inherit the rigidity of the primary CR. Due to these effects, typical calculations based on Eq.~(\ref{eq:Xeq}) are accurate only at the ten percent level or so~\cite{Ptuskin1996,SS98}.

In addition, various propagation effects cause energy change for CR rigidities $\R\lsim10$ GV. Striking examples of such effects are the latitude dependent geomagnetic cutoff of the earth and the time dependent (charge dependent and independent) solar modulation~\cite{kulbook,longbook}. Other mechanisms likely operate in interstellar space. Since the theoretical understanding of these effects is poor, we limit our analysis to CR rigidities $\R\gsim10$ GV.

The plan of this paper is as follows. In Section~\ref{s:nuc} we study the high energy radioactive nuclei data analyzed by~\cite{WS98}. We present the data in a new way, which makes the existence of an underlying time scale in the problem -- the CR residence time -- manifest. Our presentation of the data demonstrates that the charge ratio analysis of~\cite{WS98} does not suffer from gross systematics. We show how a combination of data from different nuclei species can be used to measure the functional form of the suppression due to decay. We show that a CR residence time which is mildly dependent or constant as a function of rigidity is consistent with the data.

In Section~\ref{s:pos} we consider the PAMELA positron measurements~\cite{Adriani:2008zr,Adriani:2010ib}. We argue that the positron flux is consistent with a secondary origin~\cite{Katz:2009yd}. Assuming secondary origin we compare the effects of cooling in positrons to the effects of decay in nuclei, extracting bounds on the cooling time of the positrons. These bounds can be used to test in detail the production mechanism of the positrons, and have implications for CR propagation time scales in general.

In Section~\ref{s:disc} we summarize and discuss our results. We point out that new measurements expected in the near future from the AMS-02 experiment~\cite{Bindi:2010zz} can significantly improve our quantitative understanding of CR propagation time scales in a model independent manner. Our simple analysis can readily be repeated for the new AMS-02 data.

In Appendix~\ref{app:mod} we give two propagation model examples to illustrate some general arguments made in the paper.

\section{Lessons from radioactive nuclei data}\label{s:nuc}

The measured effect of radioactive decay in nuclei receives different interpretations, depending on the adopted propagation model. At low energies $E\lsim$ 400 MeV/nuc, measurements of the isotopic ratios $^{10}$Be/$^{9}$Be, $^{26}$Al/$^{27}$Al, $^{36}$Cl/Cl and $^{54}$Mn/Mn yield, within a leaky box model interpretation, a residence time $\te\sim$ 15 Myr~\cite{Yanasak:2001}. Using the same low energy data, diffusion models yield a residence time $\te\sim100$ Myr~\cite{Strong:2007nh}. Since the isotopic ratios are only known at sub-GeV energies, they give no direct information about the energy dependence of the residence time and of the suppression of the flux due to decay.

In contrast, the decaying charge to decayed charge ratios Be/B, Al/Mg, Cl/Ar and Mn/Fe have been measured up to energies $E\sim16$ GeV/nuc (rigidities $\R\sim40$ GV~ \cite{Engelmann:1990zz}). Ref.~\cite{WS98} interpreted these charge ratios using three different models: a leaky box model, a diffusion model and a diffusion-based Monte-Carlo calculation. From the Monte-Carlo analysis, a CR residence time $\te\sim$20-30 Myr was deduced. Given that the usual leaky box and homogeneous diffusion models assume $\te\propto\X$ and that $\X\sim\R^{-0.5}$, it is interesting to note that~\cite{WS98} quotes a single value for the CR residence time. Since the data spans about a decade in rigidity, the deduced value of $\te$ would have changed by a factor of three in this range had the naive expectation been applicable.

The charge ratios Be/B, Al/Mg, Cl/Ar and Mn/Fe are predicted in a leaky box model to vary by about 20\%, 30\%, 50\% and 40\%, respectively,  between no decay and complete decay. While this estimate may not apply in general, it makes clear that systematic uncertainties are of major concern~\cite{Yanasak:2001,Mewaldt:2001}.
Measurement errors were discussed in~\cite{WS98}. Here we emphasize three types of systematic uncertainties which are most relevant for our analysis. First, uncertainties in the fragmentation cross sections are of order ten percent~\cite{Webber:2003}. Second, a potential source of uncertainty is the existence of some primary component in the $^{26}$Al, $^{36}$Cl and $^{54}$Mn fluxes~\cite{SC98Al26,SDC98Cl36}. Third, as noted in the introduction, the use of the CR grammage relation, Eq.~(\ref{eq:Xeq}), introduces additional theoretical inaccuracy of order ten percent. All three types of bias are difficult to quantify precisely. A dedicated analysis, beyond the scope of this paper, would be required to do so. In the rest of this section we show that despite this caveat, the results of~\cite{WS98} contain clear physical information, which can be used to constrain the CR residence time directly, in a model independent manner.

Following~\cite{Katz:2009yd}, we define the suppression factor due to decay of a radioactive nucleus specie $i$ as
\beq\label{eq:fs} f_{s,i}=\frac{J_i}{\frac{c}{4\pi}\,\tilde Q_i\,\X}.\eeq
The suppression factor $f_{s,i}$ can be derived from the surviving fraction $\tilde f_i$ reported in~\cite{WS98}, using the formula
\beq\label{eq:fft}f_{s,i}=\frac{\tilde f_i}{1+\frac{\sigma_i}{m_p}\X\left(1-\tilde f_i\right)},\eeq
where $\sigma_i$ is the total fragmentation cross section per ISM nucleon. Using the suppression factor $f_{s,i}$ rather than the surviving fraction $\tilde f_i$ allows us to (i) isolate the effect of decay from that of spallation losses and (ii) later on, compare directly the effect of decay of radioactive nuclei to the effect of energy losses of positrons.

\begin{figure}%\hspace{-1.5cm}
\begin{center}
\includegraphics[width=10cm]{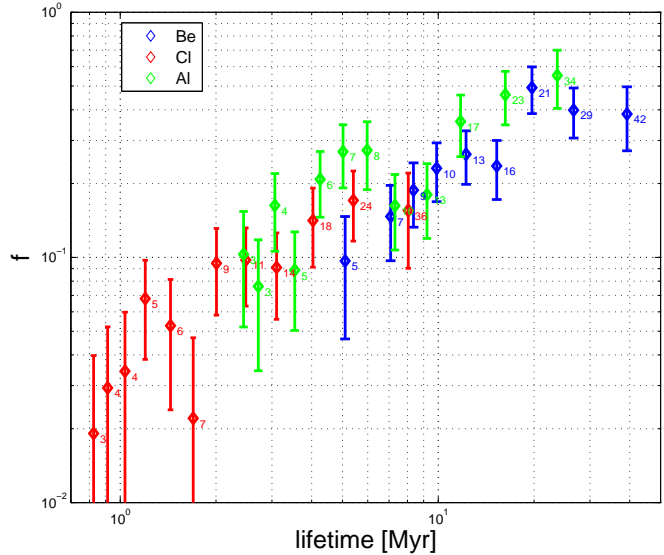}%\hfill
\end{center}
\caption{$f_{s,i}$ vs. observer frame lifetime. Numbers near data points denote the corresponding rigidities in GV.
} \label{fig:fnuc}
\end{figure}

It is instructive to examine the suppression factor due to decay vs. observer frame lifetime. This is done in Fig.~\ref{fig:fnuc} for $^{10}$Be, $^{26}$Al and $^{36}$Cl, using the fragmentation cross sections and rest frame lifetimes from Table~\ref{tab:iso} and the CR grammage from~\cite{WML2003}. We omit $^{54}$Mn, the lifetime of which is uncertain~\cite{Yanasak:2001}, and adopt a 15\% uncertainty estimate for the CR grammage and the fragmentation cross sections. A number next to each data point denotes the rigidity at that point in GV. We make the following observations: (i) The suppression factor for all three nuclei species exhibits a clear correlation with lifetime. (ii) Where available at the same lifetime, $f_{s,{\rm ^{10}Be}}\approx f_{s,{\rm ^{36}Cl}}$ and $f_{s,{\rm ^{26}Al}}\approx f_{s,{\rm ^{36}Cl}}$ to better than a factor of two.

Since the rest frame lifetimes of $^{10}$Be and $^{36}$Cl ($^{26}$Al and $^{36}$Cl) are different by a factor of five (three), data points of equal lifetime are separated by similar factors in rigidity. The approximate continuity of $f_{s,i}$ across nuclei specie thus hints that the rigidity dependence of the residence time is not strong. To assess this point quantitatively, we make some rather general assumptions.

\begin{table}[h]\begin{center}
\begin{tabular}{|c||c|c|}
  \hline
  reaction & $\tau$~[Myr] & $\sigma$~[mb] \\ \hline\hline
  $^{10}_4{\rm Be}\,\to\,^{10}_5{\rm B}$ & 2.18\,(0.09) & 210 \\
  $^{26}_{13}{\rm Al}\,\to\,^{26}_{12}{\rm Mg}$ & 1.31\,(0.06) & 410 \\
  $^{36}_{17}{\rm Cl}\,\to\,^{36}_{18}{\rm Ar}$ & 0.443\,(0.003) & 520 \\
%  $^{54}_{25}{\rm Mn}\,\to\,^{54}_{26}{\rm Fe}$ & 0.713\,(0.009)* & 685 \\
\hline\hline
\end{tabular}\caption{Decay lifetimes and fragmentation cross sections for unstable radioactive isotopes. Decay lifetimes refer to pure $\beta$ decay channels~\cite{Yanasak:2001,Donato:2001eq}. Fragmentation cross sections refer to spallation on H. We adopt a 15\% uncertainty estimate for the cross sections and correct them for an ISM composition of 90\%H+10\%He, using the prescription in~\cite{Ferrando:1988tw}.}\label{tab:iso}\end{center}\end{table}

It is natural to expect the residence time $\te$ to be a function of rigidity,
\beq\label{eq:teR}\te=\te(\R),\eeq
and $f_{s,i}$ to be a function
of $(t_i/\te)$ where $t_i$ is the observer frame lifetime. Limiting the discussion to high energy CRs with $\R>10$ GV, the data spans about a decade in lifetime. In this range, we assume an approximate power law form
\beq\label{eq:fa} f_{s,i}\approx \left(\frac{t_i}{\te}\right)^{\alpha}.\eeq
Two comments are in order: (i) In general, $\alpha$ depends on $\left(t_i/\te\right)$.\footnote{In the limit $t_i\gg\te$ we must have $\alpha\to0$. Otherwise when $t_i<\te$ we must have $\alpha>0$. Hence, $\alpha$ decreases with increasing $\left(t_i/\te\right)$.} In writing Eq.~(\ref{eq:fa}), we assume that this dependence is mild. We will see that this assumption agrees with the data. In the future, with data sampled more densely and extending to higher energy, the method we use can be straightforwardly extended to account for varying $\alpha$. (ii) The precise physical meaning of $\te$ appearing in Eq.~(\ref{eq:fa}) is model dependent. What is relevant for our discussion is that Eq.~(\ref{eq:fa}) is applicable to any nuclei specie (and later on, to positrons) with the same time scale $\te$. In Appendix~\ref{app:mod} we demonstrate how Eq.~(\ref{eq:fa}) arises in specific models.

\begin{figure}\hspace{-1.5cm}
%\begin{center}
\includegraphics[width=9cm]{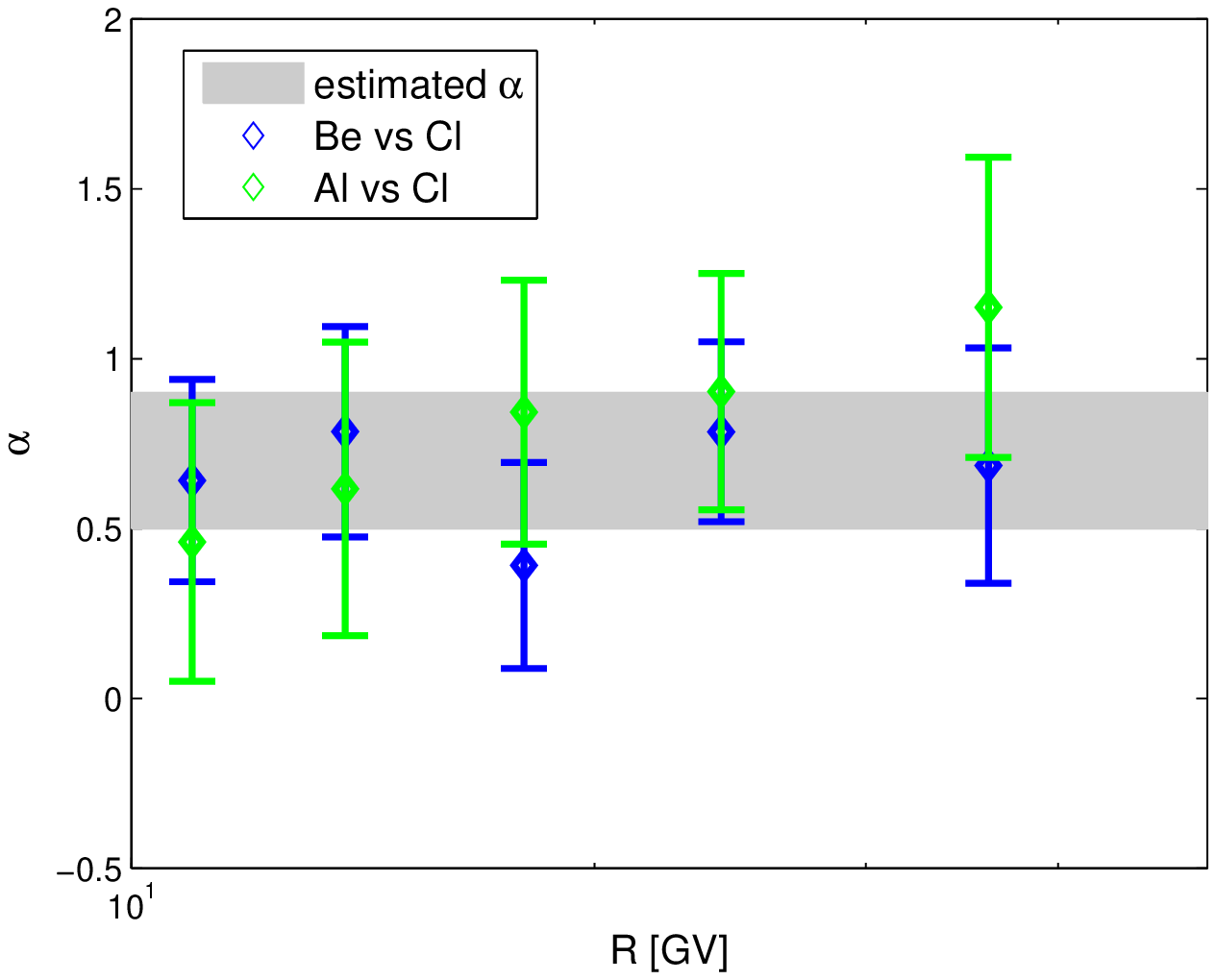}\hfill
\includegraphics[width=9cm]{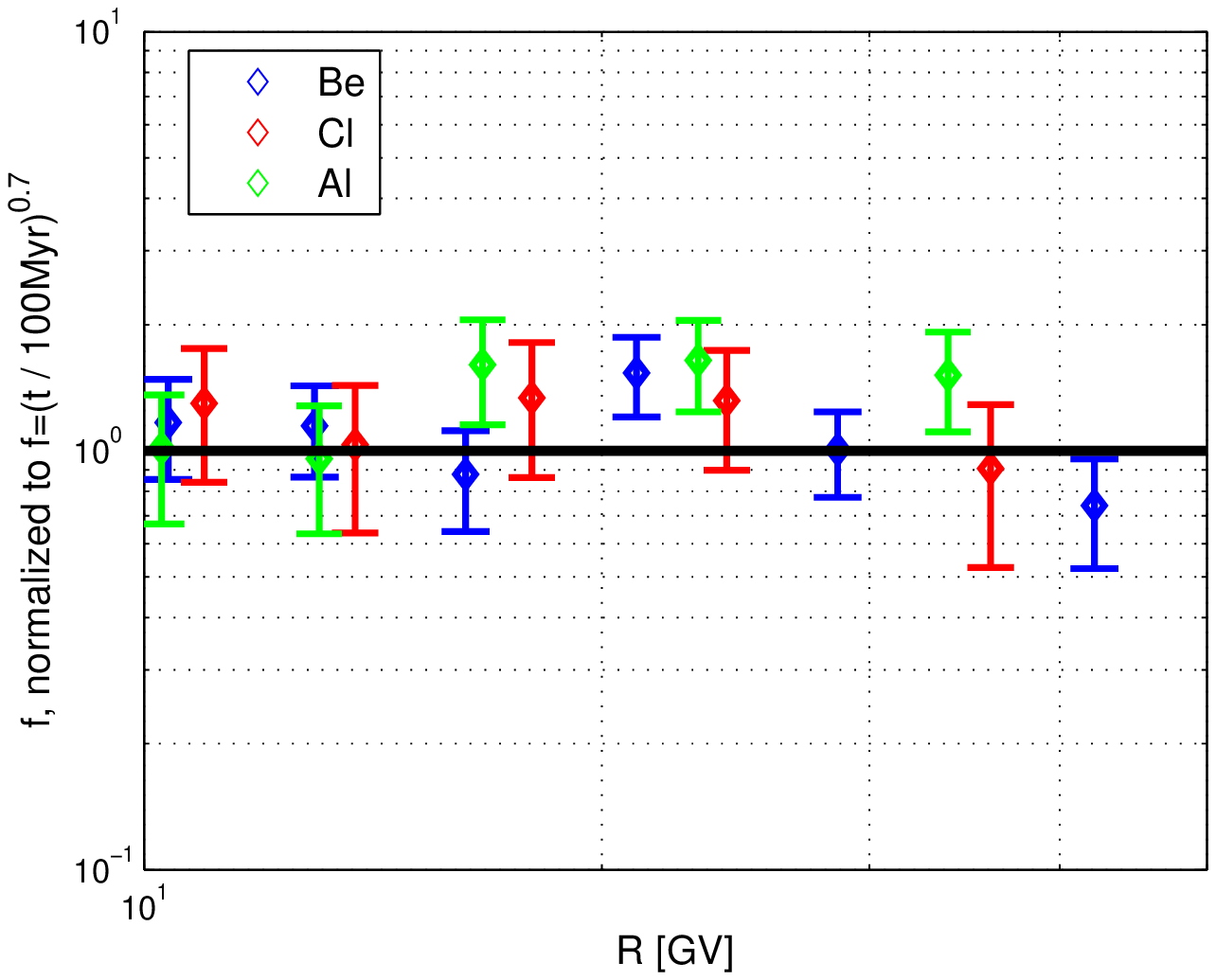}%{fR1.eps}
%\end{center}
\caption{Left: Values of $\alpha$ extracted from pairs of $f_{s,{\rm ^{10}Be}},f_{s,{\rm ^{36}Cl}}$ and $f_{s,{\rm ^{26}Al}},f_{s,{\rm ^{36}Cl}}$ measured at approximately the same rigidity.
Right: The residual rigidity dependence of $f_{s,i}$ normalized to $f_{s,i}=(t_i/\te)^\alpha$, with $\te=100\,{\rm Myr}$ and $\alpha=0.7$.
} \label{fig:fres}
\end{figure}

Under assumptions Eqs.~(\ref{eq:teR}-\ref{eq:fa}), comparing the suppression factor for two relativistic nuclei $i$ and $j$ at the same rigidity $\R'$ we have
\beq\label{eq:alpha}\log\left(\frac{f_{s,i}\left(\R'\right)}{f_{s,j}\left(\R'\right)}\right)\approx\alpha\,\log\left(\frac{A_j\,Z_i\,\tau_i}{A_i\,Z_j\,\tau_j}\right).\eeq
Thus we can extract $\alpha$, regardless of the particular form of $\te$.
A determination of $\alpha$ from Eq.~(\ref{eq:alpha}) requires nuclei with significantly
different rest frame lifetimes, since $\Delta\alpha\propto1/\log\left(\tau_i/\tau_j\right)$. In the left panel of Fig.~\ref{fig:fres} we plot values of $\alpha$, obtained by applying Eq.~(\ref{eq:alpha}) to pairs of $f_{\rm ^{10}Be},f_{\rm ^{36}Cl}$ and $f_{\rm ^{26}Al},f_{\rm ^{36}Cl}$ measured at approximately the same rigidity. The result is consistent with constant $\alpha$. Assuming constant $\alpha$ we combine the ten data pairs at $\R>10$ GV, obtaining
\beq\label{eq:a}\alpha=0.7\pm0.2.\eeq
The error estimate in Eq.~(\ref{eq:a}) is twice that arising from the errors in $f_{s,i}$ and in the nuclei lifetimes. The inferred value of $\alpha$ is shown as a gray band in the left panel of Fig.~\ref{fig:fres}.

Using Eq.~(\ref{eq:a}) and Fig.~\ref{fig:fnuc} we estimate $\te\sim$ 100 Myr in the range $\R\sim$ 10-40 GV. To make contact with the correlation depicted in Fig.~\ref{fig:fnuc}, comparing the suppression factors for two relativistic nuclei $i$ and $j$ at the same observer frame lifetime $t'$ we have
\beq\label{eq:ff}\frac{f_{s,i}\left(t'\right)}{f_{s,j}\left(t'\right)}\approx\left(\frac{\te\left(\R_j\right)}{\te\left(\R_i\right)}\right)^\alpha,\eeq
where
\beq\frac{e\,\R_i}{m_p\,c^2}=\frac{t'\,A_i}{\tau_i\,Z_i},
\;\;\;\R_j=\left(\frac{A_j\,Z_i\,\tau_i}{A_i\,Z_j\,\tau_j}\right)\,\R_i.\eeq
Given that $\R_{^{36}{\rm Cl}}\approx4\,\R_{^{10}{\rm Be}}\approx3\,\R_{^{26}{\rm Al}}$ and the preferred range $\alpha>0.4$, we verify that Fig.~\ref{fig:fnuc} constrains the rigidity dependence of the residence time. In particular, assuming a power law form
\beq\label{eq:teanzats}\te=t_{\rm esc,0}\left(\frac{\R}{10\,\rm GV}\right)^\delta,\eeq
we can extract $\alpha\delta$ from
\beq\label{eq:alphadelta}\log\left(\frac{f_{s,i}\left(t'\right)}{f_{s,j}\left(t'\right)}\right)\approx\alpha\delta\,\log\left(\frac{A_j\,Z_i\,\tau_i}{A_i\,Z_j\,\tau_j}\right).\eeq
The two pairs of $f_{^{10}{\rm Be}},f_{^{36}{\rm Cl}}$ and $f_{^{26}{\rm Al}},f_{^{36}{\rm Cl}}$ data points at minimal rigidities $\sim9$ and 10 GV give
\beq\label{eq:ad1}\alpha\delta=0\pm0.6\;\;\;\;\;(\R\approx10\,{\rm GV}),
\eeq
were the error estimate is twice that arising from the errors in $f_{s,i}$ and in the nuclei lifetimes.

In the right panel of Fig.~\ref{fig:fres} we plot the residual rigidity dependence of $f_{s,i}$ normalized to $f_{s,i}=\left(t_i/100\,{\rm Myr}\right)^{0.7}$, inferred from Eq.~(\ref{eq:a}) and Fig.~\ref{fig:fnuc}. We find no significant residual bias. Thus, Fig.~\ref{fig:fres} contains the main result of this section: the high energy charge ratio measurements are adequately described by Eq.~(\ref{eq:fa}), implying nontrivial constraints on $\alpha$ and consistent with a rigidity independent CR residence time. We make some final comments as follows:
\begin{itemize}
\item As seen from Eq.~(\ref{eq:alpha}), $\alpha$ can be determined independently of the residence time from the normalization of $f_{s,i}$ for different nuclei species. There are five Cl data points with $\R>10$ GV and about twice that number of Be and Al points with comparable rigidities. Therefore, the current data gives a fair determination of $\alpha\sim0.7$.
\item In principle, once $\alpha$ is determined, the residence time would also be measured. In practice, since for some data point $\te\sim t_i/f_{s,i}^{1/\alpha}$, the measurement of $\te$ is sensitive to the lower range of $\alpha$. Because of the limited range in rigidity, $\delta$ and $t_{\rm esc,0}$ are correlated. Thus large values of $t_{\rm esc,0}$ and correspondingly large negative $\delta$ are still allowed in conjunction with small values of $\alpha$.
\item Negative values of $\delta<-1$ are in some tension with observations. This is relevant to the interpretation of the PAMELA positron data, discussed in the next section. If $\delta<-1$, the positron suppression factor can rise with energy even under the simplest assumption of constant radiative losses.
\item $\alpha$ is significantly larger than zero at least up to $t_i\sim30$ Myr and $\R\sim30$ GV. Thus, AMS-02 data~\cite{Bindi:2010zz} (in particular Cl data) should be useful for improving the measurement of $\delta$ and $t_{\rm esc,0}$.
\item Of course, assuming Eqs.~(\ref{eq:fa}) and (\ref{eq:teanzats}) we could simply fit for $\alpha,\,\delta$ and $t_{\rm esc,0}$. However, before the role of systematic errors is clarified by a careful analysis, the precise statistical significance of such a fit is unclear and we postpone it to a future study.
\end{itemize}

For illustration, in Appendix~\ref{app:mod} we contrast the data with two specific propagation models, a diffusion model and a uniform density model (the latter is a generalization of the usual leaky box model). For $\R>$ 10 GV the results for these models in terms of $\alpha,\delta$ and $t_{\rm esc,0}$ are
\beqa\label{eq:modfit}
\alpha&\approx&0.7,\;\delta\approx0,\;t_{\rm esc,0}\approx2.7\times35\,{\rm Myr}\;\;\;({\rm leaky\, box}),\nn\\
\alpha&\approx&0.5,\;\delta\approx-0.3,\;t_{\rm esc,0}\approx300\,{\rm Myr}\;\;\;({\rm diffusion}).
\eeqa

%%%%%%%%%%%%%%%%%%%%%%%%%%%%%%%%%%%%%%%%%%%%%%%%%%%%%%%%%%%%%%%%%
%%%%%%%%%%%%%%%%%%%%%%%%%%%%%%%%%%%%%%%%%%%%%%%%%%%%%%%%%%%%%%%%%
%%%%%%%%%%%%%%%%%%%%%%%%%%%%%%%%%%%%%%%%%%%%%%%%%%%%%%%%%%%%%%%%%
\section{Lessons from the PAMELA positron data}\label{s:pos}

Recently, the flux of cosmic ray antiprotons~\cite{Boezio:2008mp,:2010rc} and
positrons~\cite{Adriani:2008zr,Adriani:2010ib} was measured by the PAMELA satellite-borne experiment.\footnote{See~\cite{Schubnell:2009gk,Adriani:2010ib} for cautionary notes regarding systematic uncertainties in the positron measurement.} While the antiproton measurements stand in accord with common diffusion models, the positron measurements do not, revealing an excess with respect to model predictions. The `positron excess' triggered numerous attempts to explain the acclaimed `non-secondary' origin of the positrons.
However, a careful examination of the positron data does not reveal any excess with respect to model independent estimates, assuming secondary production~\cite{Katz:2009yd}. Below we analyze the implications of the measured positron fraction under the secondary production hypothesis. We also briefly reiterate the main points of~\cite{Katz:2009yd}.

A main point we wish to convey is that the PAMELA positron measurement is, first and foremost, a measurement of the radiative cooling. To see this, note that it is possible to test the secondary production spectrum of the positrons indirectly, by examining the flux of another secondary CR which does not suffer propagation energy losses and which originates from the same interactions. The first candidate for such a test is the antiproton flux.

\begin{figure}[hbp]\hspace{-1.5cm}
%\begin{center}
\includegraphics[width=9cm]{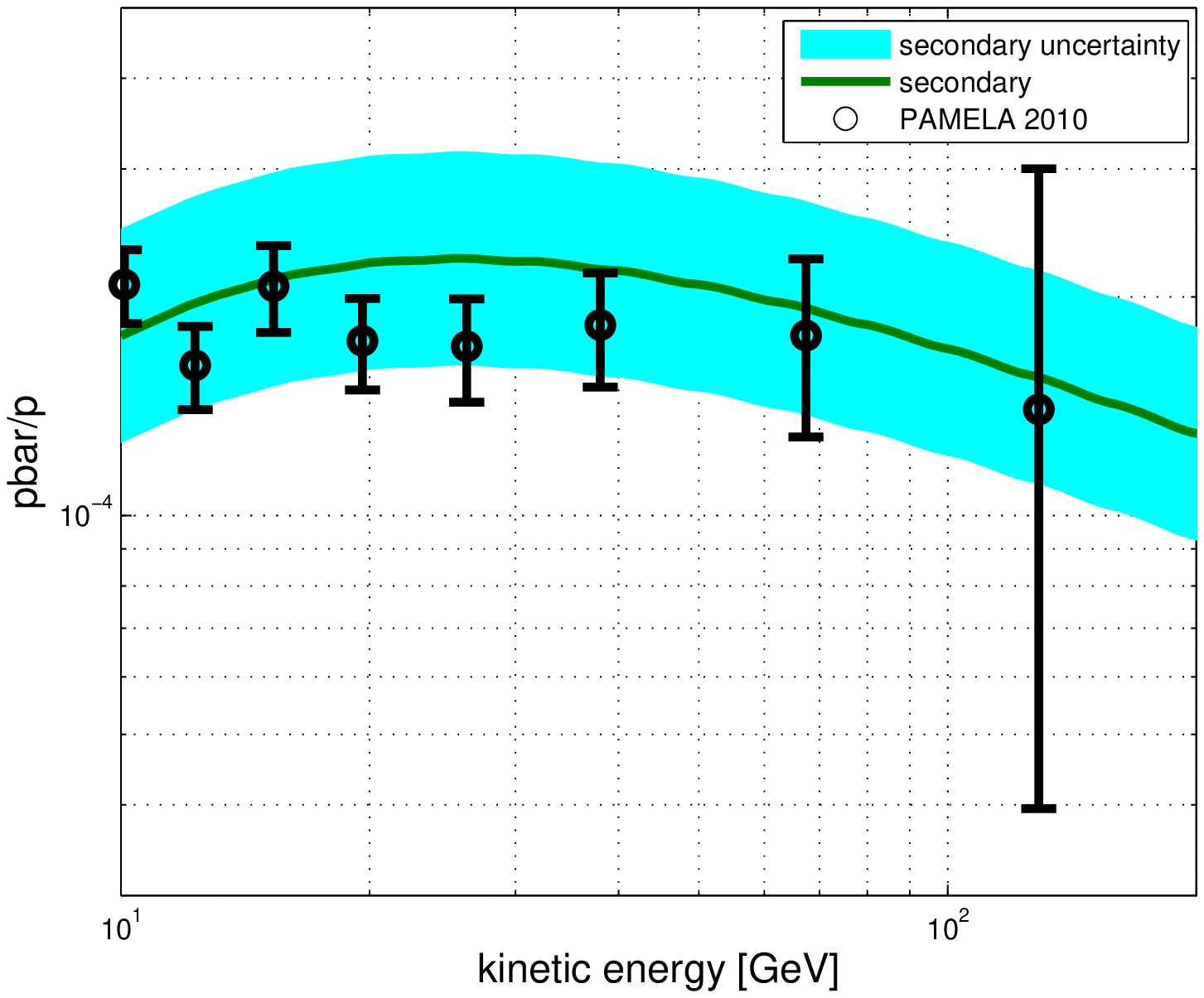}\hfill
\includegraphics[width=9cm]{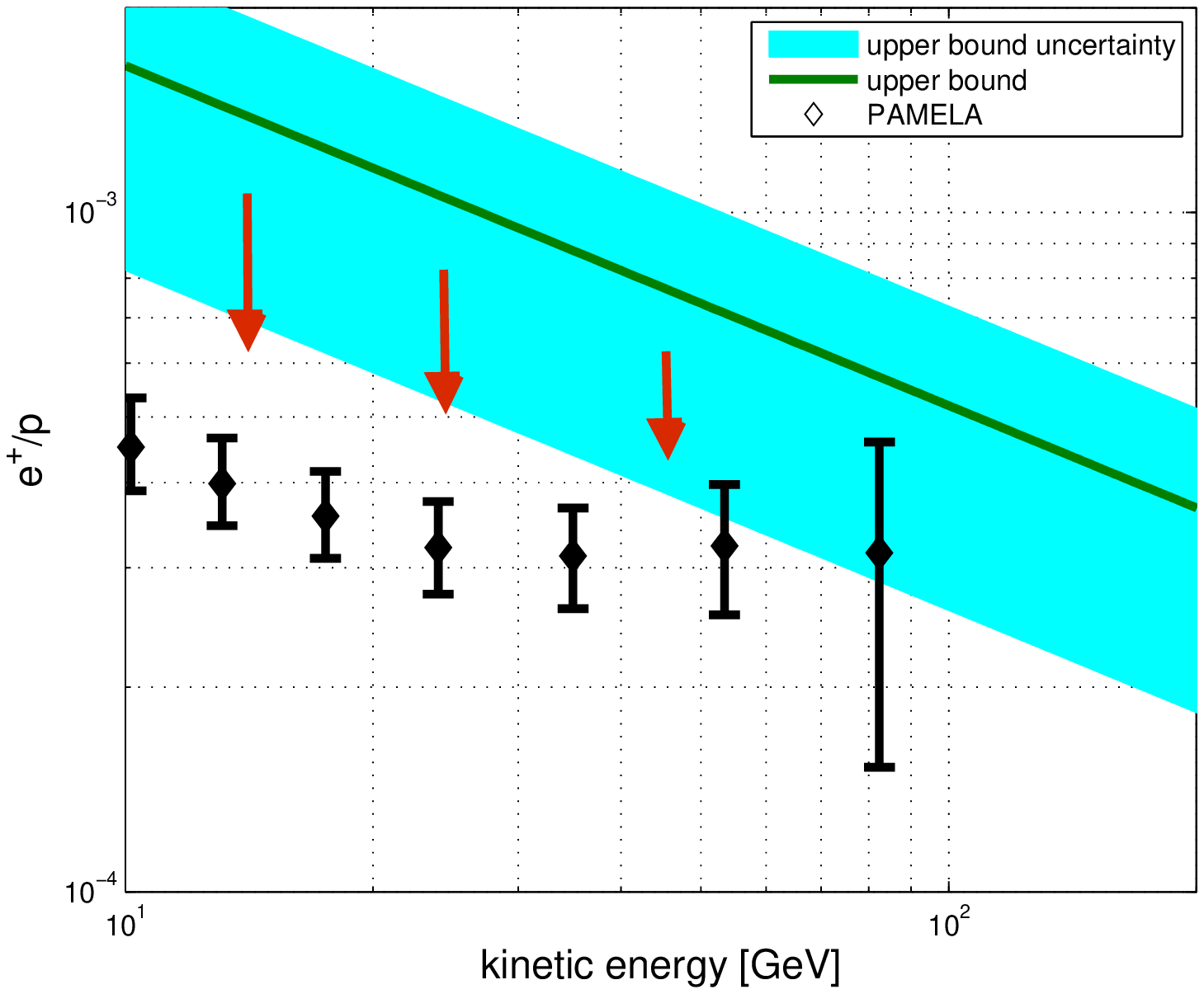}
%\end{center}
\caption{Left: antiproton to proton flux ratio. Right: positron to proton flux ratio.} \label{fig:pospbar}
\end{figure}

In the left panel of Fig.~\ref{fig:pospbar} we show the results of applying Eq.~(\ref{eq:Xeq}) to the calculation of the antiproton to proton flux ratio. The details of the calculation are given in~\cite{Katz:2009yd}. We show a 40\% error estimate, representative of relevant cross section uncertainties, as well as the latest PAMELA results~\cite{:2010rc}.
Having verified the applicability of Eq.~(\ref{eq:Xeq}) for antiprotons, we can now find the cooling suppression for positrons by computing the positron flux from Eq.~(\ref{eq:Xeq}), neglecting radiative losses, and comparing the result to the observed flux.

To put the positron and antiproton analyses on equal grounds we give the flux ratio of positrons to their leading progenitors, protons. No published PAMELA results currently exist for the individual $e^+,e^-,e^\pm$ or $p$ fluxes. Extracting the positron to proton ratio from the reported $e^+/e^\pm$ fraction~\cite{Adriani:2010ib} requires combining measurements from different experiments. Here we perform this combination aiming to illustrate a representation of the data which -- unlike the $e^+/e^\pm$ fraction -- naturally lends itself to theoretical interpretation. A sounder experimental procedure will be preferable once complete data are published by the same experiment. We interpolate the total electronic flux from FERMI~\cite{collaboration:2010ij}, available down to $\R=7$ GV, using solar modulation parameter $\Phi=500$ MeV. The proton flux we use is that of~\cite{Moskalenko:2001ya}, for which we adopt a 10\% normalization error.

The result is shown in the right panel of Fig.~\ref{fig:pospbar}. The error band around the theoretical upper bound is explained below. As indicated by the arrows, PAMELA measured the cooling suppression of the positrons.

Next, we quantify the effect of losses. The positron cooling suppression factor can be defined in direct analogy to the suppression due to decay in radioactive nuclei. We find
\beq\label{eq:fc} f_{s,e^+}=\frac{J_{e^+}}{\frac{c}{4\pi}\,\tilde Q_{e^+}\,\X
}\approx0.6\times10^{3}\left(\frac{\R}{10\,\rm GV}\right)^{0.5}\times\frac{J_{e^+}(\R)}{J_p(\R)},\eeq
for the cross section parameterization of~\cite{Tan:1984ha}. Using the cross sections from~\cite{Kamae:2006bf} yields a factor of $\sim2$ increase in the value of $f_{s,e^+}$, compared to (\ref{eq:fc}), for $\R\lsim100$ GV~\cite{Delahaye:2008ua}. This factor of two ambiguity is responsible for the main uncertainty in the derivation of the upper bound. Considering this ambiguity we adopt a factor of two error estimate, shown in Fig.~\ref{fig:pospbar}, for the theoretical calculation of the positron to proton ratio. Other uncertainties of order 20\% arise e.g. from the effect of heavy nuclei in the primary CR flux and in the ISM~\cite{Mori:2009te}.

The values of $f_{s,e^+}$ for positrons and $f_{s,i}$ for some radioactive nucleus should be comparable, if examined at the particular rigidity, wherein the corresponding cooling time and observer frame lifetime are similar. This observation applies if the cooling time can be approximated by a power law,
\beq t_c\propto\R^{-\delta_c},\eeq
and is explained in~\cite{Katz:2009yd} by noting that decay of radioactive nuclei and energy losses of positrons are represented by formally similar terms in a general transport equation. Thus, we expect
\beq\label{eq:fspos} f_{s,e^+}\approx\left(\frac{t_c}{\te}\right)^\alpha,\eeq
where $\alpha$ and $\te$ are the same as in Eq.~(\ref{eq:fa}). Eq.~(\ref{eq:fspos}) is accurate up to corrections of order $(\gamma-1-\delta_c)^{-\alpha}$, where $\gamma\sim3$ is the observed spectral index of the positrons. Using $\alpha\sim0.7$, in accordance with Sec.~\ref{s:nuc}, these corrections are at the level of 10\% for $\delta_c\sim1$ (40\% for $\delta_c\sim0$).

The positron cooling time can be estimated by
\beq\label{eq:tc}\tc\approx10\,{\rm Myr}\,\left(\frac{\R}{30\,{\rm
GV}}\right)^{-1}\left(\frac{\bar U_T}{1\,{\rm eV\,cm^{-3}}}\right)^{-1},\eeq
where $\bar U_T$ is the total electromagnetic energy density in the propagation region, averaged over CR histories. Comparing $f_{s,i}$ of some nucleus with $f_{s,e^+}$ of positrons at the same rigidity $\R'$ we expect
\beq\label{eq:rt}\frac{f_{s,i}(\R')}{f_{s,e^+}(\R')}
\approx\left[\left(\frac{\tau_i}{1.5\,{\rm Myr}}\right)\,\left(\frac{\R'}{20\,{\rm
GV}}\right)^2\,\left(\frac{\bar U_T}{1\,{\rm eV\,cm^{-3}}}\right)\right]^\alpha.\eeq
If $\bar U_T$ is rigidity independent, then the ratio Eq.~(\ref{eq:rt}) should rise as $f_{s,i}/f_{s,e^+}\sim\R^{2\alpha}$. Note that a rigidity dependent $\bar U_T$ is possible. In particular, if $\bar U_T$ decreases with increasing CR rigidity, then the ratio Eq.~(\ref{eq:rt}) could have a milder rise, $f_{s,i}/f_{s,e^+}\propto\R^{\alpha(1+\delta_c)}$ with $\delta_c<1$.\footnote{For $\sim$ 50 GV positrons, Klein-Nishina corrections imply that the starlight component in $U_T$ induces $\delta_c<1$. Without extreme assumptions regarding the starlight intensity~\cite{Stawarz:2009ig}, these corrections are of the scale of current uncertainties and we defer their discussion to a more refined treatment.}

In the rest of this section, we use Eqs.~(\ref{eq:fspos}-\ref{eq:rt}) to analyze the implications of having radioactive nuclei data in conjunction with positron data in the same range in rigidity. First, we show that the measured positron flux is consistent quantitatively with secondary production. Then, we assume a secondary origin for the positrons and derive constraints on the mean traversed electromagnetic energy density, $\bar U_T$. Finally, we comment on the energy dependence of the cooling suppression factor of the positrons.

\begin{figure}[hbp]
\begin{center}
\includegraphics[width=10cm]{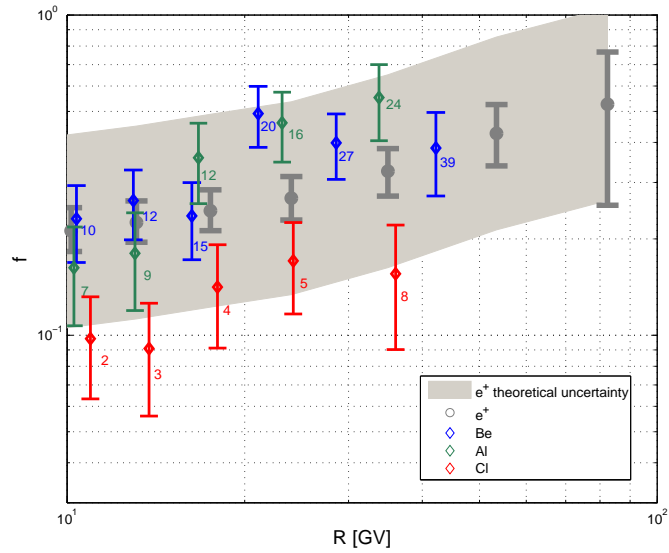}
\end{center}
\caption{The suppression factor due to cooling ($e^+$) and radioactive decay (nuclei). Numbers near nuclei data points denote the lifetime at that point in Myr.
}
\label{fig:fRpos}
\end{figure}

Using Eq.~(\ref{eq:fc}) we plot $f_{s,e^+}$ vs. rigidity in Fig.~\ref{fig:fRpos}, together with the suppression factors due to decay for $^{10}$Be, $^{26}$Al and $^{36}$Cl nuclei. Nuclei lifetimes in Myr are specified next to each data point. The shaded band denotes the theoretical uncertainty. Error bars show error estimates from the experimental procedure.

Fig.~\ref{fig:fRpos} is in broad agreement with the observation made in~\cite{Katz:2009yd}, namely that the cooling suppression factor measured by PAMELA for positrons is consistent quantitatively with that found for radioactive nuclei. This observation makes stronger the case for a secondary origin for the positrons. In particular, in~\cite{Katz:2009yd} it was pointed out that $f_{s,e^+}\sim f_{s,\rm ^{10}Be}$ at $\R\sim20$ GV, in agreement with Eq.~(\ref{eq:rt}) for plausible $\bar U_T\sim$ 1 eV/cm$^3$. We comment that while the central data do not show a clear rising pattern of $f_{s,i}/f_{s,e^+}$ with rigidity, $f_{s,i}/f_{s,e^+}\sim\R$ or so (with, e.g., $\alpha\approx0.5$ and constant $\bar U_T$) is perfectly possible within the calculation uncertainties and given the visible scatter.

Assuming secondary positrons, we can use Eq.~(\ref{eq:rt}) together with Fig.~\ref{fig:fRpos} to infer the value of $\bar U_T$. First, assuming that the PAMELA result is not grossly wrong, we can set $f_{s,e^+}=f_{s,^{36}\rm Cl}$ to derive an upper bound
\beq\label{eq:Ulim}\bar U_T<5\,\left(\frac{\R}{20\,{\rm
GV}}\right)^{-2}\,{\rm eV\,cm^{-3}}\;\;\;\;\;({\rm e^+\,and\,Cl}),\eeq
valid in the range $\R\sim$ 10-40 GV. The bound Eq.~(\ref{eq:Ulim}) is nontrivial: for $\R=40$ GV it implies $\bar U_T<1.25$ eV/cm$^3$. For reference, the local value of $U_T$ is determined from CMB, IR and optical surveys~\cite{Porter:2005qx} to be $U_{T,\rm disc}=\mathcal{O}(1)$ eV/cm$^3$.
Second, Be and Al data give potentially stronger, though less robust, constraints
\beqa\label{eq:bealU}\bar U_T&\sim&\left(0.5-3\right)\,\left(\frac{\R}{20\,{\rm
GV}}\right)^{-2}\,{\rm eV\,cm^{-3}}\;\;\;\;\;({\rm e^+,\,Be\,and\,Al}),
\eeqa
valid in the same range as Eq.~(\ref{eq:Ulim}).

Note that $\bar U_T<U_{CMB}\approx0.25$ eV/cm$^3$ cannot occur physically. If, using Eq.~(\ref{eq:rt}) with upcoming experimental data, such value will be reliably deduced for $\bar U_T$ it will falsify the assumption of a secondary origin for the positrons. From Fig.~\ref{fig:fRpos} we see that Cl data at $\sim50$ GeV/nuc will be very useful for such a test.

Finally we comment on the implications of flat or mildly rising $f_{s,e^+}$, as derived in Fig.~\ref{fig:fRpos} from the various data~\cite{Adriani:2010ib,collaboration:2010ij,Moskalenko:2001ya}. By Eq.~(\ref{eq:fspos}), the positron suppression factor should rise if $\te$ drops faster than $t_c$ with increasing rigidity~\cite{Katz:2009yd}. The only information which directly constrains $\te\propto\R^\delta$ is the radioactive nuclei data analyzed in Section~\ref{s:nuc}. This analysis shows that values of $\delta<-1$ are in some tension with observations. However, improved charge ratio measurements, in particular Cl/Ar data extending to $\R\sim$ 100 GV (and so to lifetime $t_{\rm Cl}\sim$ 40 Myr, comparable to the currently existing Be/B measurements), will be required to conclusively rule out (or measure) $\delta<-1$. Regarding the cooling time $t_c\propto\R^{-\delta_c}$, effective $\delta_c<1$ arise if Klein-Nishina corrections to Eq.~(\ref{eq:tc}) are important, or if $\bar U_T$ decreases with increasing rigidity. A rigidity dependent $\bar U_T$ is observationally and theoretically conceivable. In fact, such behavior of $\bar U_T$ will resemble the decreasing traversed matter density inferred from stable secondary to primary nuclei flux ratios. To conclude, at present we cannot yet rule out the possibility that $\delta_c+\delta<0$, for which $f_{s,e^+}$ rising with rigidity occurs with secondary production. In the next section we summarize methods by which the secondary origin hypothesis for the positrons can be challenged.

\section{Summary and discussion}\label{s:disc}

In this paper we presented a model independent analysis of the propagation time scales of cosmic ray (CR) nuclei and positrons. Motivated by the prospects for improved measurements in the near future~\cite{Adriani:2008zr,Ahn:2008my,collaboration:2010ij,Guzik:2008zz,Bindi:2010zz}, our analysis demonstrated that it is possible to extract significant quantitative information from CR measurements under general assumptions, without committing to any particular propagation model.

In Sec.~\ref{s:nuc} we studied the radioactive nuclei charge ratio measurements, first analyzed in~\cite{WS98}. Examining the suppression of the flux due to decay vs. observer frame lifetime, the clear correlation across nuclei specie, seen in Fig.~\ref{fig:fnuc}, is indicative that the charge ratio analysis is not plagued with large systematics. (See, in contrast,~\cite{Yanasak:2001,Mewaldt:2001}.) In addition, the effects of a possible primary component in the $^{26}$Al or $^{36}$Cl flux appear to be small on the scale of the measurement errors. This makes a combined analysis of the charge ratios a useful model independent tool, first in measuring the effect of decay and, second, in constraining the CR residence time. Using charge ratio data at $\R\sim$ 10-40 GV, we showed that a rigidity independent residence time is consistent with observations, and that the suppression of the flux for all nuclei species is adequately described by $f_{s,i}=\left(t_i/\te\right)^{\alpha}$, with $\te=100\,{\rm Myr}$ and $\alpha=0.7$.

In Sec.~\ref{s:pos} we analyzed the PAMELA positron data~\cite{Adriani:2008zr,Adriani:2010ib}. We argued that the positron flux is consistent with a secondary origin~\cite{Katz:2009yd}. Assuming secondary positrons, we derived an upper bound on the mean electromagnetic energy density traversed by the positrons, $\bar U_T<1.25$ eV/cm$^3$ at a rigidity of $\R=$ 40 GV. This bound is close to the locally measured electromagnetic energy density~\cite{Porter:2005qx}. A stronger, though less robust bound can be inferred from Be and Al measurements, and may indicate that the positrons spend significant fraction of their confinement time in regions far from the Galactic disc, where the electromagnetic energy density is lower than the locally observed value.

Methods by which the secondary origin hypothesis for the positrons can be challenged are summarized as follows:
\begin{enumerate}
\item A simple first test for secondary positrons is the condition \beq\label{eq:fl1}f_{s,e^+}<1.\eeq This test is independent of radioactive nuclei data. As seen in Fig.~\ref{fig:pospbar}, this condition is satisfied by current data~\cite{Katz:2009yd}.
\item A second, and potentially stronger test for the origin of the positrons is the requirement \beq\label{eq:UUCMB}\bar U_T>U_{CMB}.\eeq This test can challenge secondary production when $f_{s,e^+}<1$, and can be performed robustly with Cl measurements. At present, our evaluation shows that $\bar U_T\sim1\,{\rm eV/cm^3}>U_{CMB}$, and so the test is passed in the range where $e^+$ and Cl data coexist.
\item Finally, improved radioactive nuclei data will yield a model independent measurement of $\alpha$, describing the functional form of the suppression of the flux due to decay or loss (see Eqs.~(\ref{eq:fa}),(\ref{eq:fspos})) and of the rigidity dependence of the residence time, $\te\propto\R^\delta$. Given such a measurement, the detailed energy dependence of the ratio $f_{s,i}/f_{s,e^+}\propto\R^{\alpha(1+\delta_c)}$, for any radioactive nucleus $i$ will further determine the energy dependence of the cooling time, $t_c\propto\R^{-\delta_c}$. If future measurements (and future PAMELA re-analyses~\cite{Adriani:2010ib}) will confirm a rising positron fraction, then a violation of the condition
    \beq\label{eq:dd1}\delta+\delta_c<0\eeq
    will establish a non-secondary origin for the positrons.
\end{enumerate}

The effect of decay of radioactive nuclei, defined in Eq.~(\ref{eq:fa}) and specified by a measurement of $\alpha$, has a deep connection with propagation. In the general class of models in which CR production occurs mainly in the thin Galactic disc, the quantity $f_{s,i}$ in the limit of strong losses is given by~\cite{ginbook}
\beq\label{eq:fdeep}f_{s,i}\sim\frac{t_i\,L_{\rm esc}}{\te\,L_i}.\eeq
Here $t_i$ is the observer frame lifetime, $L_i$ is the scale height above the disc occupied by the decaying nuclei and $\te$ is the escape time from the propagation region of scale height $L_{\rm esc}$. $\te$ and $L_{\rm esc}$ are defined for stable CRs, and would apply to the specie $i$ had its lifetime been infinite~\cite{Ptuskin:1998}. The functional form of $f_{s,i}$ is determined by the nature of the CR motion in the trapping volume. Diffusion models have $L\sim\sqrt{t}$ and give $f_{s,i}\sim\sqrt{t_i/\te}$ or $\alpha\sim0.5$. Leaky box models impose $L_{\rm esc}=L_i$, giving $f_{s,i}\sim t_i/\te$ or $\alpha\sim1$.\footnote{As shown in Appendix~\ref{app:mod}, for the leaky box model the strong loss result ceases to apply for $t_i/\te>\mathcal{O}(0.1)$, leading to $\alpha\sim0.7<1$. The strong loss result $\alpha\sim1$ could in principle be tested using short-lived nuclei such as $^{14}_6C$.} Models of compound diffusion~\cite{ginbook} predict $L\sim t^{0.25}$ and thus $f_{s,i}\sim (t_i/\te)^{0.75}$ or $\alpha\sim0.75$.
It is worth pointing out in this context that the currently fashionable homogeneous diffusion models with large halo and thin disc predict $\alpha=0.5$ to high accuracy, see Appendix~\ref{app:mod}. This value of $\alpha$ is currently allowed by data. Establishing $\alpha>0.5$ will rule out this class of models.

The AMS-02 experiment~\cite{Bindi:2010zz} has the capabilities to take the analysis to a new level of accuracy in the near future.\footnote{We note that in~\cite{Pato:2010ih} the implications of future AMS-02 measurements were analyzed in the context of a diffusion model.} AMS-02 will resolve CR charge all the way to $Z=26$ and up to TeV/nuc. In addition, AMS-02 will measure isotopic ratios including $^9$Be/$^{10}$Be up to 10 GeV/nuc~\cite{Arruda:2005tw}. This measurement will be a valuable check of the interpretation of the charge ratios. Finally, AMS-02 will also measure the antiproton and positron fluxes up to $\sim$300 GeV~\cite{Bindi:2010zz}, assessing the validity of the PAMELA measurements~\cite{Boezio:2008mp,:2010rc,Adriani:2008zr,Adriani:2010ib} and putting the combined analysis of positrons and radioactive nuclei on firmer observational grounds.

Further progress which is potentially within immediate reach will follow a publication of the individual spectra of $e^+,e^-,p$ and nuclei by the PAMELA collaboration. These data can be used to extract quantities such as $e^+/p$ and $e^+/\bar p$, which are theoretically clean probes of CR propagation, in a direct manner.

\section*{Acknowledgments}

We thank C.~Delaunay for useful discussions and Y.~Hochberg, B.~Katz and especially E.~Waxman for  comments on the manuscript.

\begin{appendix}
\section{Propagation model examples}\label{app:mod}

\begin{figure}%\hspace{-0.5cm}
\includegraphics[width=9cm]{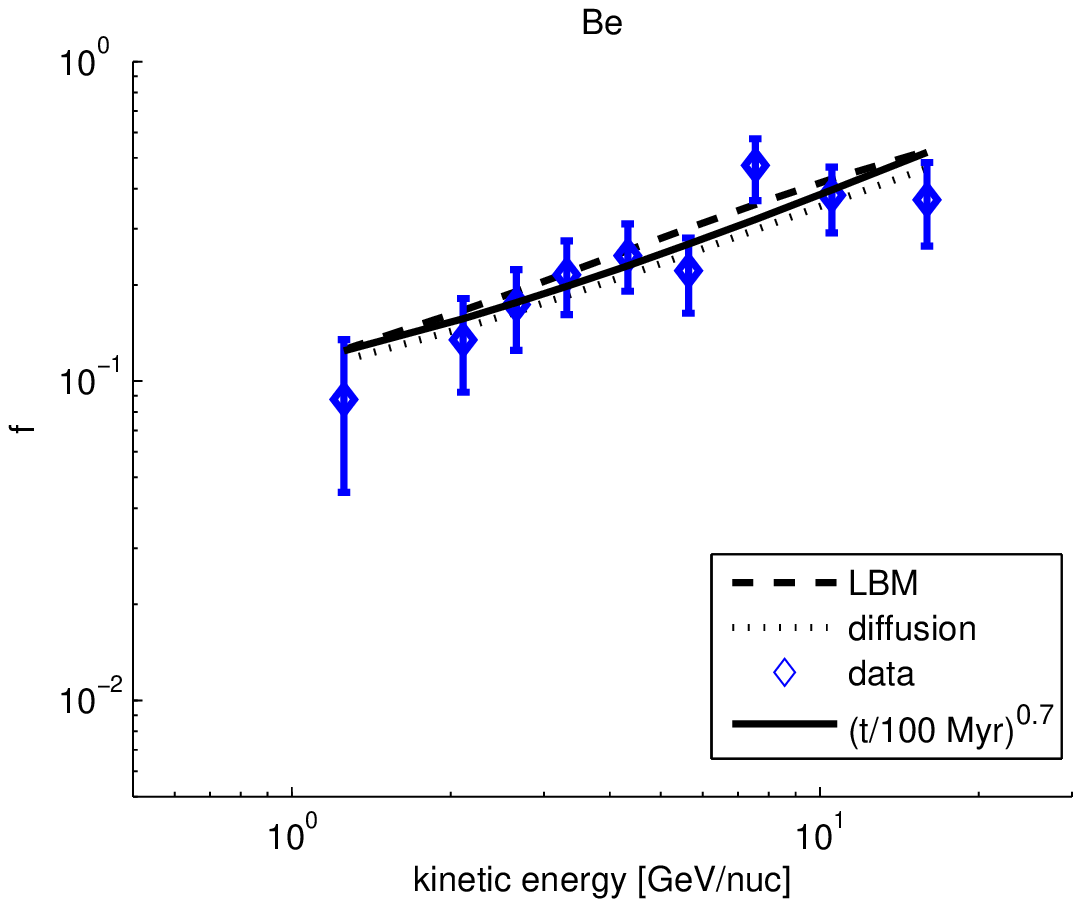}\hfill
\includegraphics[width=9cm]{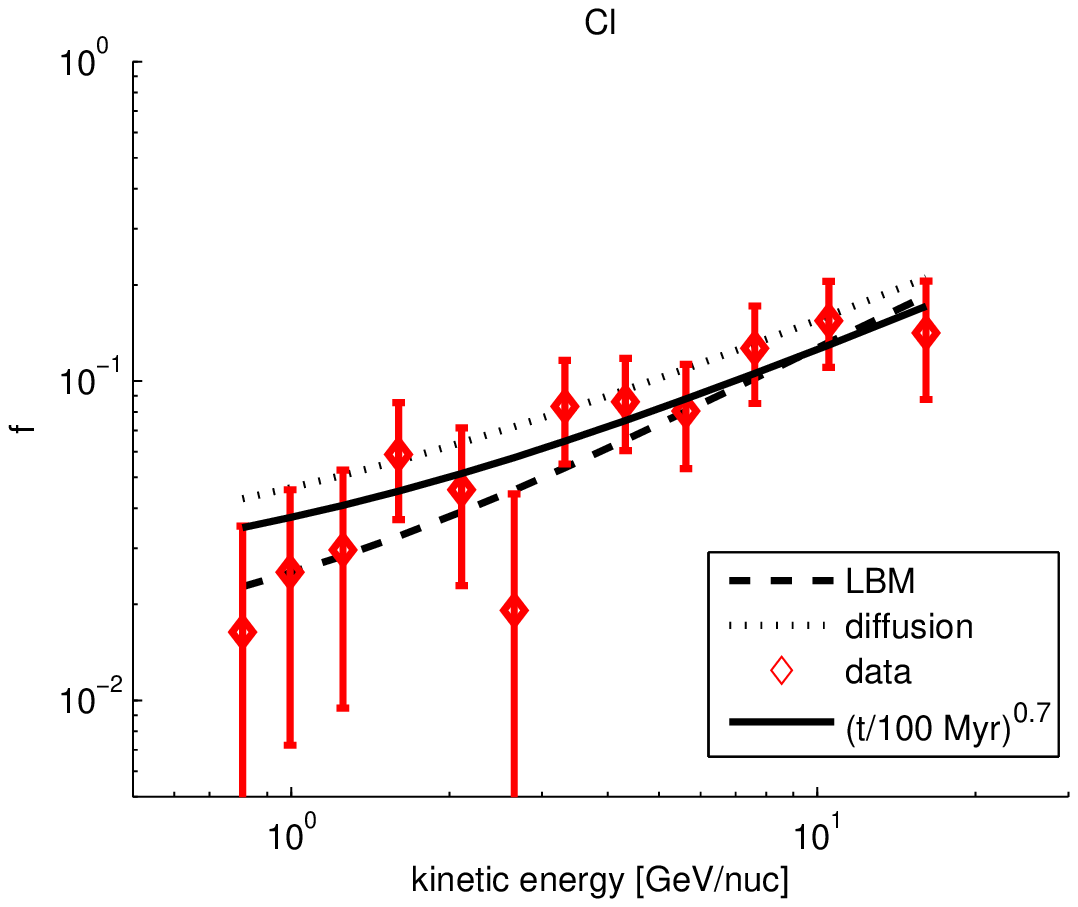}\\
\includegraphics[width=9cm]{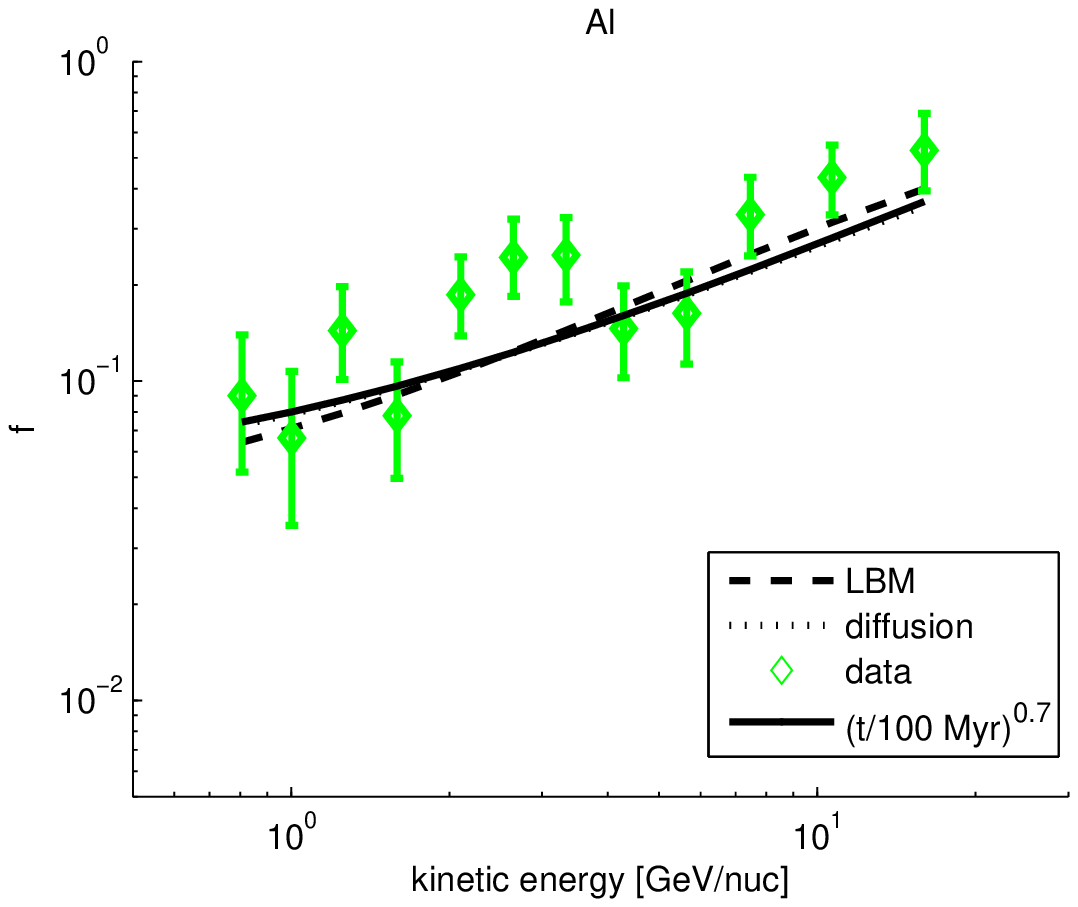}\hfill
\includegraphics[width=9cm]{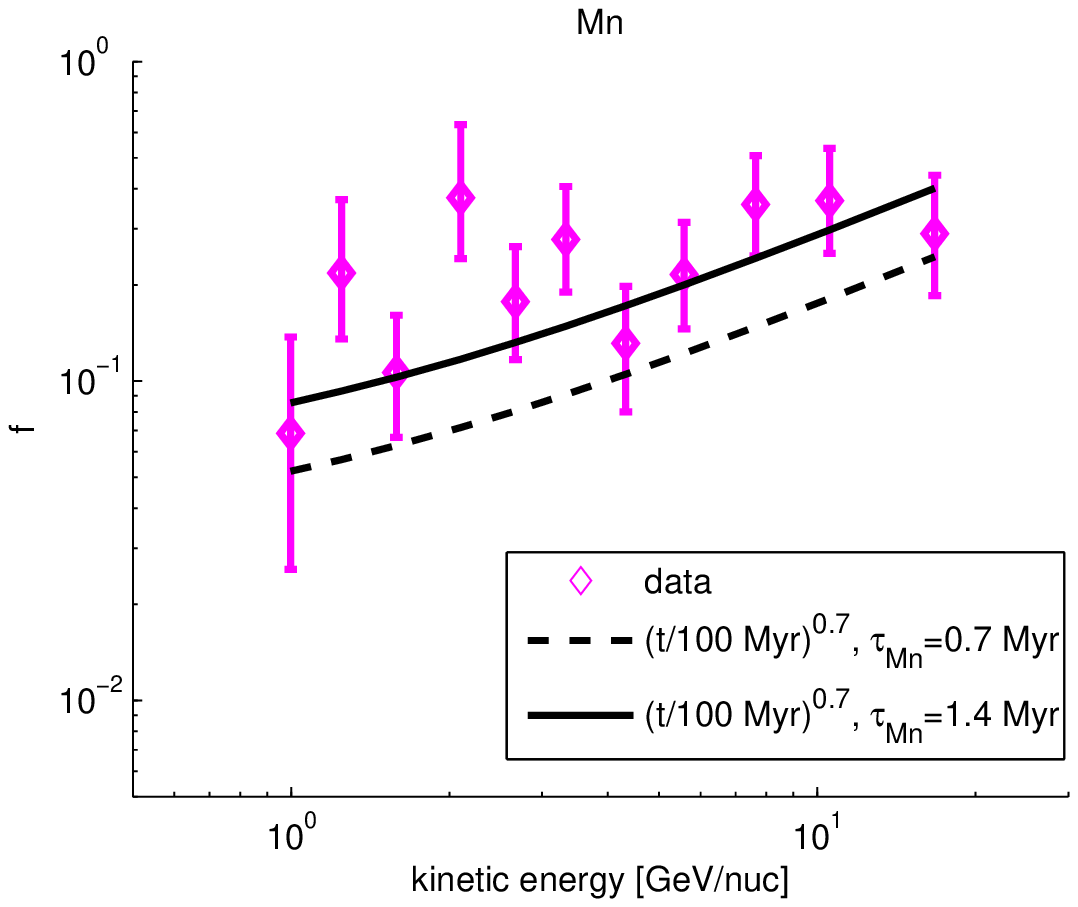}
\caption{The suppression factor due to decay $f_{s,i}$. For Be, Al and Cl, smooth, dashed and dotted lines denote the empirical result of Sec.~\ref{s:nuc}, the leaky box model (LBM) and the diffusion model solutions, respectively. Here we extend the plot to low energy $\R<10$ GV. We also plot Mn, the data of which did not take part in the fit, showing the empirical result applied to two different assumptions for the decay lifetime.}\label{fig:radnuc}
\end{figure}

Here we demonstrate the implications of the radioactive nuclei data for two specific models. First we consider a version of the leaky box model, with spatially uniform CR distribution at a given rigidity but with a rigidity dependent halo scale height $L=L(\R)$. Second we consider thin disc diffusion models with a large halo. The models predict:
\beqa\label{eq:fsmod}
f_{s,i}&=&\frac{1}{1+\te/t_i},\;\;{\rm leaky\,box},\nn\\
f_{s,i}&=&\sqrt{t_i/\te}\,\tanh\left(\sqrt{\te/t_i}\right),\;\;{\rm diffusion}. \eeqa
We fit the suppression factor for $^{10}$Be, $^{26}$Al and $^{36}$Cl using Eqs.~(\ref{eq:fsmod}) and~(\ref{eq:teanzats}), so that our fit parameters are $t_{\rm esc,0}$ and $\delta$. In each case we arrange the parameters of the model (halo size and escape time for the leaky box, and halo size and diffusion coefficient for the diffusion model) such as to reproduce the measured CR grammage~\cite{WML2003}. The results for both models are shown in Fig.~\ref{fig:radnuc}. We make the following comments:
\begin{itemize}
\item The suppression factor $f_{s,i}$ in Eq.~(\ref{eq:fsmod}) involves the nucleus specie label $i$ only via the lifetime $t_i$. Any explicit dependence on the fragmentation cross sections and CR grammage is canceled. This would not have been the case had we worked with the surviving fraction $\tilde f_i$ (as easily verified from Eq.~(\ref{eq:fft})). We expect this result to hold in general for disc+halo models, in which the CR composition in the regions where the secondaries are produced is similar to the CR composition near earth.
\item For $\R>10$ GV, values of $f_{s,\rm ^{10}Be}$, $f_{s,\rm ^{26}Al}$ and $f_{s,\rm ^{36}Cl}$ obtained from Eq.~(\ref{eq:fsmod}) satisfy the power law assumption Eq.~(\ref{eq:fa}) to better than 10\%, with $\alpha=$0.7 (0.5) for the leaky box (diffusion) cases. For the leaky box model, the parameter $t_{\rm esc}$ has the meaning of $2.7$ times the escape time of the CRs. For the diffusion model, $t_{\rm esc}$ is the average time it takes a CR produced on the disc to reach  the halo boundary.
\item Both the leaky box and diffusion models accommodate the data reasonably well.  We find \beqa\te&\approx&2.7\times\left(30-45\,\right)\times\left(\R/{10\,\rm GV}\right)^{0\pm0.2}\,{\rm Myr},\;\;\;{\rm
leaky\,box,}\nn\\
\te&\approx&\left(200-400\,\right)\times\left(\R/{10\,\rm GV}\right)^{-0.3\pm0.3}\,{\rm
Myr},\;\;\;{\rm diffusion}.\nn\eeqa
\end{itemize}

\end{appendix}

%% **********************************************************************

\end{document}